\documentclass{elsart}%
\usepackage{amsfonts}
\usepackage{amsmath}
\usepackage{amssymb}
\usepackage{graphicx}%
\begin{document}
\begin{frontmatter}%
\title{Space-time dimensionality $\mathcal{D}%
$ as complex variable: calculating loop integrals using dimensional
recurrence relation and analytical properties with respect to $\mathcal{D}$.}%

\author{R.N. Lee}%
\ead{r.n.lee@inp.nsk.su}

\address
{The Budker Institute of Nuclear Physics and Novosibirsk State University, 630090, Novosibirsk, Russia}%

\begin{abstract}
We show that dimensional recurrence relation and analytical properties of the loop integrals
as functions of complex variable $\mathcal{D}$ (space-time dimensionality) provide a regular
way to derive analytical representations of loop integrals. The representations derived have a
form of exponentially
converging sums. Several examples of the developed technique are given.%
\end{abstract}%


\end{frontmatter}%

\section{Introduction}

High-order radiative corrections to different amplitudes and cross section
become nowadays more and more important both in QED and QCD. In QED this is
due to the high precision of the modern spectroscopy experiments, while in QCD
the radiative corrections play important role due to the strength of the
interaction. The radiative corrections are expressed in terms of the loop
integrals, so the possibility to calculate the latter is also very important.
Several powerful techniques have been developed for the calculation of the
multiloop integrals. Among them is the IBP (integration-by-part) reduction
procedure \cite{Tkachov1981,ChetTka1981}, which allows one to reduce arbitrary
loop integral to some finite set of master integrals. Owing to IBP reduction,
the problem of calculation of loop integrals is reduced to the calculation of
master integrals. It is important to understand that IBP reduction also helps
in the calculation of the master integrals. Indeed, differentiating the master
integral with respect to some external invariant or mass, we obtain linear
combination of integrals of the same topology or its subtopologies. Applying
IBP reduction procedure, we can express this combination via the same set of
master integrals. Acting in this way, one can obtain the system of ordinary
differential equations for all master integrals \cite{Kotikov1991,Remiddi1997}%
. The general solution of this system depends on several constants
parametrizing the solution of the corresponding homogeneous system. These
constants can be fixed from the value of the integrals at some specific
kinematic parameters, or from their asymptotics in which one invariant becomes
big, or small, in comparison with others. This method can be applied to the
integrals with several external parameters.

However, the integrals with one scale are also important for the applications.
For them, the method of differential equations does not work and one has to
rely on other methods. In addition to the direct calculation using, e.g.,
Mellin-Barnes representation, one can try the method of difference equations.
One of the variants of the difference equation is described by Laporta in Ref.
\cite{Laporta2000}. In this method one considers the generalized master
integral which is formed by raising one of the massive denominators of the
master integral of interest to arbitrary power $x$. Using IBP reduction one
can obtain the difference equation with respect to $x$. The part,
corresponding to the general solution of the homogeneous equation is then
fixed from the large-$x$ asymptotics. The resulting expression for the master
integral has the form of factorial series with power-like (harmonic)
convergence (the general term of the series falls down as $k^{-\alpha}$, where
$k$ is the summation variable, $\alpha>1$). This method has been successfully
applied for several multiloop tadpole integrals, see, e.g., Refs.
\cite{SchrVuo2005,BejdSch2006}. Due to the power-like convergence of sums, the
difficulties of obtaining high-precision results rapidly grow with the number
of digits. The multi-digit results, on the other hand, are important for the
\texttt{pslq} algorithm \cite{FergBai1991}, which allows one to restore the
analytical result from the high-precision numerical one, provided the basis of
transcedental numbers is known.

Another variant of the difference equation is the dimensional recurrence
relation connecting the master integrals in different dimensions, suggested by
Tarasov in Ref. \cite{Tarasov1996}. We remind, that, in contrast to the
differential equation, where the solution of the homogeneous equation is
parametrized by several arbitrary constants, in the case of difference
equation the corresponding solution is parametrized by several arbitrary
periodic functions. In Refs. \cite{Tarasov1996,Tarasov2000} the
large-$\mathcal{D}$ asymptotics has been used for fixing these functions. This
asymptotics has been derived from the explicit parametric form of the
integrals. However, for multiloop integrals the calculation of this
asymptotics appears to be hardly accessible. This is why the method of the
dimensional recurrence relation was mainly applied to the calculation of
one-loop integrals. It is worth noting that for the case of several external
invariants, one may rely on the combination of differential and difference
equations methods, see Ref. \cite{Tarasov2006}.

Recently, in Ref. \cite{KiriLee2009} the method of the dimensional recurrence
relation was successfully applied to the calculation of certain four-loop
tadpole. In that paper, an arbitrary periodic function parametrizing the
solution of the homogeneous equation was numerically shown to be equal to zero
by using the Laporta-like difference equation with respect to the power of
massive denominator. However this vanishing of the homogeneous part was rather
an exception than a rule, and in many known cases the homogeneous part is not
zero. Needless to say that in these cases the correct choice of the periodic
functions not only can be hardly justified, but plainly difficult to guess. On
the other hand, all examples of application of the dimensional recurrence
relation show that results obtained with the help of this method have a form
of exponentially converging series (the general term of the series falls down
as $a^{-k}$, where $k$ is the summation variable, $a>1$). The rapid
convergence of these series allows one to use the \texttt{pslq} algorithm to
express their $\epsilon$-expansion in terms of conventional transcedental
numbers, like multiple zeta-functions.

In this paper, we describe a method of the calculation of loop integrals based
on the usage of Tarasov's dimensional recurrence relation. The key point of
our approach is the use of the analytical properties of the integrals as
functions of the complex variable $\mathcal{D}$. These properties allow one to
fix the periodic functions up to some constants which can be found from the
calculation of the integral at some definite $\mathcal{D}$.

The paper is organized as follows. In the next Section, we introduce notations
and formulate a general path of finding the master integrals. In Section III
we briefly analyse the analytical properties of the parametric representation
needed for our consideration. In Section IV we give several examples of the
application of the formulated approach. Section V contains discussions of the
generalization of the formulated approach to the topologies with more than one
master integral. The last Section gives short conclusion.

\section{Dimensional recurrence relation}

Assume that we are interested in the calculation of the $L$-loop integral
depending on $E$ linearly independent external momenta $p_{1},\ldots,p_{E}$.
There are $N=L(L+1)/2+LE$ scalar products depending on the loop momenta
$l_{i}$:%

\begin{equation}
s_{ik}=l_{i}\cdot q_{k}\,,\quad i=1,\ldots,L,\quad k=1,\ldots,L+E,
\end{equation}
where $q_{1,\ldots,L}=l_{1,\ldots,L}$, $q_{L+1,\ldots,L+E}=p_{1,\ldots,E}$.

The loop integral has the form%
\begin{equation}
J^{\left(  \mathcal{D}\right)  }\left(  \boldsymbol{n}\right)
=\int\frac{d^{\mathcal{D}}l_{1}\ldots d^{\mathcal{D}}l_{L}}{\pi^{L\mathcal{D}/2}D_{1}^{n_{1}%
}D_{2}^{n_{2}}\ldots D_{N}^{n_{N}}} \label{eq:J}%
\end{equation}
where the scalar functions $D_{\alpha}$ are linear polynomials with respect to
$s_{ij}$. The functions $D_{\alpha}$ are assumed to be linearly independent
and to form a complete basis in the sense that any non-zero linear combination
of them depends on the loop momenta, and any $s_{ik}$ can be expressed in
terms of $D_{\alpha}$. For further purposes it is convenient to introduce the
operators $A_{\alpha}$ and $B_{\alpha}$, see Ref. \cite{Lee2008}, acting as
follows%
\begin{align}
\left(  A_{i}J^{\left(  \mathcal{D}\right)  }\right)  \left(  n_{1}%
,\ldots,n_{N}\right)   &  =n_{i}J^{\left(  \mathcal{D}\right)  }\left(
n_{1},\ldots,n_{i}+1,\ldots,n_{N}\right)  ,\nonumber\\
\left(  B_{i}J^{\left(  \mathcal{D}\right)  }\right)  \left(  n_{1}%
,\ldots,n_{N}\right)   &  =J^{\left(  \mathcal{D}\right)  }\left(
n_{1},\ldots,n_{i}-1,\ldots,n_{N}\right)  .
\end{align}

Relation between the loop integrals in different dimensions has been first
introduced in Ref. \cite{Tarasov1996}. It was derived using the parametric
representation for the loop integrals. For the integral given by some graph,
the result of Ref. \cite{Tarasov1996} may be represented as follows%
\begin{equation}
J^{\left(  \mathcal{D}-2\right)  }\left(  \mathbf{n}\right)  =\mu^L\sum
_{\text{trees}}\left(  A_{i_{1}}\ldots A_{i_{L}%
}J^{\left(  \mathcal{D}\right)  }\right)  \left(  \mathbf{n}\right)
,\label{eq:Raising}%
\end{equation}
where $i_{1},\ldots,i_{L}$ numerate the chords of the tree, and $\mu=\pm 1$ for the
euclidean/pseudoeuclidean case, respectively. It is natural to label this relation as a raising
one since it expresses any integral in $\mathcal{D}-2$ dimensions via the integrals in
$\mathcal{D}$ dimensions. It may be also convenient to use the lowering dimensional relation
which we will obtain now. Our derivation is based on the Baikov's approach \cite{Baikov1997}
in which one passes from the integration over loop momenta to the integration
over scalar products . The master formula can be represented as%
\begin{align}
\int\frac{d^{\mathcal{D}}l_{1}\ldots d^{\mathcal{D}}l_{L}}{\pi^{L\mathcal{D}%
/2}}\,f &  =\frac{\mu^L\pi^{-LE/2-L\left(  L-1\right)  /4}}{\Gamma\left[  \left(
\mathcal{D}-E-L+1\right)  /2,\ldots,\left(  \mathcal{D}-E\right)  /2\right]
}\nonumber\\
&  \times\int\left(  \prod_{i=1}^{L}\prod_{j=i}^{L+E}ds_{ij}\right)
\frac{\left[  V\left(  l_{2},\ldots l_{L},p_{1},\ldots,p_{E}\right)  \right]
^{\left(  \mathcal{D}-E-L-1\right)  /2}}{\left[  V\left(  p_{1},\ldots
,p_{E}\right)  \right]  ^{\left(  \mathcal{D}-E-1\right)  /2}}%
f,\label{eq:Baikov}%
\end{align}
where%
\begin{equation}
V\left(  q_{1},\ldots,q_{M}\right)  =%
\begin{vmatrix}
q_{1}^{2} & \dots & q_{1}\cdot q_{M}\\
\vdots & \ddots & \vdots\\
q_{1}\cdot q_{M} & \dots & q_{M}^{2}%
\end{vmatrix}
\end{equation}
is a Gram determinant constructed on the vectors $q_{1},\ldots,q_{M}$ and $f$
is arbitrary function of scalar products $s_{ij}$. For our purposes, of
course, we choose $f=\left[  D_{1}^{n_{1}}D_{2}^{n_{2}}\ldots D_{N}^{n_{N}%
}\right]  ^{-1}$. Due to the aforementioned properties of basis $\left\{
D_{1},\ldots,D_{N}\right\}  $, the function $V\left(  l_{2},\ldots l_{L}%
,p_{1},\ldots,p_{E}\right)  $ has the form of some polynomial of degree $L+E$
of $D_{1},\ldots,D_{N}$ (and of external invariants, which is not important
for our purposes):%
\begin{equation}
V\left(  l_{2},\ldots l_{L},p_{1},\ldots,p_{E}\right)  =P\left(  D_{1}%
,D_{2},\ldots,D_{N}\right)
\end{equation}
Replacing $\mathcal{D}\rightarrow\mathcal{D}+2$ in Eq. (\ref{eq:Baikov}), we
obtain the lowering dimensional recurrence relation%
\begin{equation}
J^{\left(  \mathcal{D}+2\right)  }\left(  \mathbf{n}\right)  =\frac
{(2\mu)^{L}\left[  V\left(  p_{1},\ldots,p_{E}\right)  \right]  ^{-1}}{\left(
\mathcal{D}-E-L+1\right)  _{L}}\left(  P\left(  B_{1},B_{2},\ldots
,B_{N}\right)  J^{\left(  \mathcal{D}\right)  }\right)  \left(  \mathbf{n}%
\right)  .\label{eq:Lowering}%
\end{equation}

Comparing Eqs.(\ref{eq:Raising}) and (\ref{eq:Lowering}) we see that in order
to obtain the former, one has to analyse the graph corresponding to the loop
integral, while the latter can be obtained without any reference to the graph.
So, for computer implementation the lowering dimensional relation can be more
convenient. On the other hand, Eq.(\ref{eq:Raising}) is more "economic" since
the right-hand side of it contains integrals with indices shifted by at most
$L$, while the right-hand side of Eq. (\ref{eq:Lowering}) may contain
integrals with indices shifted by $L+E$. Of course, after IBP reduction, both
raising and lowering recurrence relations are equivalent.

In case when the integral in the left-hand side of Eq. (\ref{eq:Raising}) or Eq.
(\ref{eq:Lowering}) is master integral, IBP reduction of the right-hand side gives a difference
equation for this master integral. If the topology contains only one master integral, the
right-hand side, after IBP reduction, contains this master integral and master integrals of the
subtopologies, which we assume to be found by the same method, or some other. In case of
several master integrals in the topology, we can obtain in the same way a system of
difference equations for them. We will concentrate below on the case of one master integral in
the topology of interest. The discussion of the case of several-master topology is given in
Section V. So, after the IBP reduction we obtain from, e.g., raising relation the following
dimensional recurrence
relation%
\begin{equation}
J^{\left(  \mathcal{D}-2\right)  }=C\left(  \mathcal{D}\right)  J^{\left(
\mathcal{D}\right)  }+R\left(  \mathcal{D}\right)  ,\label{eq:DRR}%
\end{equation}
where $R\left(  \mathcal{D}\right)  $ is non-homogeneous part constructed of
the master integrals of subtopologies in $\mathcal{D}$ dimensions. The
coefficient $C\left(  \mathcal{D}\right)  $ is a rational function of
$\mathcal{D}$ which can always be represented as%
\begin{equation}
C\left(  \mathcal{D}\right)  =c\frac{\prod_{i}\left(  a_{i}-\mathcal{D}%
/2\right)  }{\prod_{j}\left(  b_{j}-\mathcal{D}/2\right)  }%
\end{equation}
where $c$ is some constant, and $a_{i}$ ($b_{i}$) are the zeros of the
numerator (denominator). We can construct the solution of the homogeneous part
of the equation as follows:%
\begin{equation}
J_{0}^{\left(  \mathcal{D}\right)  }=\omega\left(  \mathcal{D}\right)
\,/\Sigma\left(  \mathcal{D}\right)  ,\label{eq:homogeneous}%
\end{equation}
where $\omega\left(  \mathcal{D}\right)  =\omega\left(  \mathcal{D}+2\right)
$ is arbitrary periodic function of $\mathcal{D}$ and $\Sigma^{-1}\left(
\mathcal{D}\right)  $ is any specific non-zero solution of the homogeneous
equation. One of possible choices of $\Sigma\left(  \mathcal{D}\right)  $ is
\begin{equation}
\Sigma^{-1}\left(  \mathcal{D}\right)  =\,c^{-\mathcal{D}/2}\frac{\prod
_{i=1}^{n}\Gamma\left(  a_{i}-\mathcal{D}/2\right)  }{\prod_{j=1}^{m}%
\Gamma\left(  b_{j}-\mathcal{D}/2\right)  }.\label{eq:integrating factor}%
\end{equation}
By analogy with the notion of integrating factor in the theory of differential
equations, we will call $\Sigma\left(  \mathcal{D}\right)  $ the \emph{summing
factor}. The choice of the summing factor is by no means unique. Given one
summing factor, we may multiply it by any periodic function of $\mathcal{D}$
to obtain another. In particular, another natural choice is%
\begin{equation}
\Sigma^{-1}\left(  \mathcal{D}\right)  =\,\left[  \left(  -1\right)
^{n+m}c\right]  ^{-\mathcal{D}/2}\frac{\prod_{j=1}^{m}\Gamma\left(
\mathcal{D}/2+1-b_{j}\right)  }{\prod_{i=1}^{n}\Gamma\left(  \mathcal{D}%
/2+1-a_{i}\right)  }.
\end{equation}
Let us now substitute $J^{\left(  \mathcal{D}\right)  }=\Sigma^{-1}\left(
\mathcal{D}\right)  g\left(  \mathcal{D}\right)  $ in Eq. (\ref{eq:DRR}).
After multiplying both parts by $\,\Sigma\left(  \mathcal{D}-2\right)  $, we
obtain
\begin{equation}
g\left(  \mathcal{D}-2\right)  =g\left(  \mathcal{D}\right)  +r\left(
\mathcal{D}\right)  ,
\end{equation}
where $r\left(  \mathcal{D}\right)  =R\left(  \mathcal{D}\right)
\Sigma\left(  \mathcal{D}-2\right)  $. Suppose now that we can decompose
$r\left(  \mathcal{D}\right)  $ into two, $r_{+}\,\left(  \mathcal{D}\right)
$ and $r_{-}\left(  \mathcal{D}\right)  $, decreasing fast enough at
$\pm\infty$, respectively:
\begin{equation}
r\left(  \mathcal{D}\right)  =r_{+}\left(  \mathcal{D}-2\right)  +r_{-}\left(
\mathcal{D}\right)  ,\quad r_{\pm}\left(  \mathcal{D}\pm2k\right)
\overset{k\rightarrow\infty}{<}a^{k},\quad\left\vert a\right\vert <1
\end{equation}
Note the shift of the argument in the definition of $r_{+}\left(
\mathcal{D}\right)  $. This decomposition also determines the corresponding
decomposition
\begin{equation}
R\left(  \mathcal{D}\right)  =R_{+}\left(  \mathcal{D}-2\right)  +R_{-}\left(
\mathcal{D}\right)  ,
\end{equation}
such that%
\begin{align}
\left\vert \lim\limits_{\mathcal{D}\rightarrow+\infty}\frac{C\left(
\mathcal{D}+2\right)  R_{+}\left(  \mathcal{D}+2\right)  }{R_{+}\left(
\mathcal{D}\right)  }\right\vert  &  <1,\nonumber\\
\left\vert \lim\limits_{\mathcal{D}\rightarrow-\infty}\frac{C^{-1}\left(
\mathcal{D}-2\right)  R_{-}\left(  \mathcal{D}-2\right)  }{R_{-}\left(
\mathcal{D}\right)  }\right\vert  &  <1.\label{eq:Rpm}%
\end{align}

The general solution reads%
\begin{align}
J^{\left(  \mathcal{D}\right)  } &  =\Sigma^{-1}\left(  \mathcal{D}\right)
\omega\left(  \mathcal{D}\right)  +\sum_{k=0}^{\infty}s_{+}\left(
\mathcal{D},k\right)  -\sum_{k=0}^{\infty}s_{-}\left(  \mathcal{D},k\right)
,\label{eq:general}\\
s_{+}\left(  \mathcal{D},k\right)   &  =\frac{r_{+}\left(  \mathcal{D}%
+2k\right)  }{\Sigma\left(  \mathcal{D}\right)  }=\left[  \left(  -1\right)
^{n+m}c\right]  ^{k}\frac{\prod\limits_{i=1}^{n}\left(  \frac{\mathcal{D}}%
{2}+1-a_{i}\right)  _{k}}{\prod\limits_{j=1}^{m}\left(  \frac{\mathcal{D}}%
{2}+1-b_{j}\right)  _{k}}R_{+}\left(  \mathcal{D}+2k\right)  ,\nonumber\\
s_{-}\left(  \mathcal{D},k\right)   &  =\frac{r_{-}\left(  \mathcal{D}%
-2k\right)  }{\Sigma\left(  \mathcal{D}\right)  }=c^{-1-k}\frac{\prod
\limits_{j=1}^{m}\left(  b_{j}-\frac{\mathcal{D}}{2}\right)  _{k+1}}%
{\prod\limits_{i=1}^{n}\left(  a_{i}-\frac{\mathcal{D}}{2}\right)  _{k+1}%
}R_{-}\left(  \mathcal{D}-2k\right)  .\nonumber
\end{align}
Note that the specific nonhomogeneous solution does not depend on the
particular choice of the summing factor. Eq.(\ref{eq:general}) clearly
demonstrates that, in order to find $J^{\left(  \mathcal{D}\right)  }$, we
have to fix the periodic function $\omega\left(  \mathcal{D}\right)  $. This
function can be conveniently understood as a function of the complex variable
\begin{equation}
z=e^{i\pi\mathcal{D}}%
\end{equation}
and below we switch the notations $\omega\left(  \mathcal{D}\right)
\rightarrow\omega\left(  z\right)  $. Understanding $\omega$ as a function of
$z$, we automatically account for the periodicity due to identity
$e^{i\pi\left(  \mathcal{D}+2\right)  }=e^{i\pi\mathcal{D}}$. Our idea of
fixing $\omega\left(  z\right)  $ is simple and yet quite efficient as we will
demonstrate in Section IV. The functions $r_{\pm}$ in the right-hand side of
Eq. (\ref{eq:general}) are assumed to be known and, in particular, their
analytical properties are known. The integral $J^{\left(  \mathcal{D}\right)
}$ in the left-hand side is not known (this is the purpose of our
investigation), but we can discover some of its analytical properties from,
e.g., its parametric representation. Note that we do not have to analyse the
integral in the whole complex plane of $\mathcal{D}$, but only in an arbitrary
stripe of width $2$, i.e., in $S=\left\{  \mathcal{D}|\quad
\,d<\operatorname{Re}\mathcal{D}\leqslant d+2\right\}  $, which we will call
the \emph{basic stripe}. The factor $\Sigma\left(  \mathcal{D}\right)  $ is
also a known analytical function of $\mathcal{D}$. This allows us to fix
$\omega\left(  z\right)  $ up to some constant which can be found from the
value of the integral $J^{\left(  \mathcal{D}\right)  }$ at one given value of
$\mathcal{D}$. There is a number of technical conveniencies in this approach.
First, as we have said above, the summing factor $\Sigma\left(  \mathcal{D}%
\right)  $ can be chosen in different forms and we are free to choose the one
with the most convenient analytical properties. Second, we can pass from using
one integral of certain topology as a master to using another integral of the
same topology as a master. Finally, we are free to choose the convenient basic
stipe in which we perform the analysis.

So, summarizing, the suggested path of finding a master integral is the
following (we assume we can do IBP reduction):

\begin{enumerate}
\item Make sure all master integrals in subtopologies are known. If it is not
so, start from calculating them.

\item Pass to a suitable master integral. It is convenient to choose a master integral which
    is finite in some interval $\mathcal{D\in}\left(  d,d+2\right)  $. For this purpose, e.g.,
    increase powers of some massive propagators.

\item Construct dimensional recurrence relation for this master integral. Due to step 1, the
    nonhomogeneous part of the recurrence relation is known.

\item Find a general solution of this recurrence relation depending on function $\omega\left(
    z\right)  $.

\item Fix the singularities of this function by analysing the analytical
properties of the master integrals and summing factor. The convenient choices
of summing factor and of basic stipe are at your disposal.

\item If needed, fix the remaining constants from the value of the integral at
some space-time dimension $\mathcal{D}$.
\end{enumerate}

The step 5 is the key point of our approach. We will present the applications
of the formulated approach on several examples, but first we would like to
demonstrate how some of the analytical properties of the integral can be
determined from its parametric representation.

\section{Parametric representation}

Let us consider the parametric representation of some $L$-loop integral in
Euclidean space with $I$ internal lines (see, e.g., Ref. \cite{Itzykson1980}):%
\begin{equation}
J^{\left(  \mathcal{D}\right)  }=\Gamma\left(  I-L\mathcal{D}/2\right)  \int
dx_{1}\ldots dx_{I}\delta\left(  1-%
{\textstyle\sum}
x_{i}\right)  \frac{\left[  Q\left(  x\right)  \right]  ^{\mathcal{D}L/2-I}%
}{\left[  P\left(  x\right)  \right]  ^{\mathcal{D}\left(  L+1\right)  /2-I}%
}.\label{eq:parametric}%
\end{equation}
The polynomials $Q\left(  x\right)  $ and $P\left(  x\right)  $ are determined
by the graph. For our consideration it is important only that both these
functions are nonnegative in the whole integration region,%
\[
Q\left(  x\right)  \geqslant0,\qquad P\left(  x\right)  \geqslant0.
\]
Suppose now that the parametric representation converges for all $\mathcal{D}$
in some interval $\left(  d_{1},d_{2}\right)  $.Then it is easy to see that
Eq. (\ref{eq:parametric}) determines $J^{\left(  \mathcal{D}\right)  }$ as a
holomorphic function on the whole stripe $S=\left\{  \mathcal{D}%
|\quad\operatorname{Re}\mathcal{D\in}\left(  d_{1},d_{2}\right)  \right\}  $.
Moreover, as $\operatorname{Im}\mathcal{D}$ tends to $\pm\infty$ while
$\operatorname{Re}\mathcal{D}=d\mathcal{\in}\left(  d_{1},d_{2}\right)  $ is
fixed, we can estimate%
\begin{align}
\left\vert J^{\left(  \mathcal{D}\right)  }\right\vert  &  =\left\vert
\Gamma\left(  I-L\mathcal{D}/2\right)  \right\vert \left\vert \int
dx_{1}\ldots dx_{I}\delta\left(  1-%
{\textstyle\sum}
x_{i}\right)  \frac{\left[  Q\left(  x\right)  \right]  ^{\mathcal{D}L/2-I}%
}{\left[  P\left(  x\right)  \right]  ^{\mathcal{D}\left(  L+1\right)  /2-I}%
}\right\vert \nonumber\\
&  \leqslant\left\vert \Gamma\left(  I-L\mathcal{D}/2\right)  \right\vert \int
dx_{1}\ldots dx_{I}\delta\left(  1-%
{\textstyle\sum}
x_{i}\right)  \left\vert \frac{\left[  Q\left(  x,P,m\right)  \right]
^{\mathcal{D}L/2-I}}{\left[  P\left(  x\right)  \right]  ^{\mathcal{D}\left(
L+1\right)  /2-I}}\right\vert \nonumber\\
&  =\left\vert \Gamma\left(  I-L\mathcal{D}/2\right)  \right\vert \int
dx_{1}\ldots dx_{I}\delta\left(  1-%
{\textstyle\sum}
x_{i}\right)  \frac{\left[  Q\left(  x,P,m\right)  \right]  ^{dL/2-I}}{\left[
P\left(  x\right)  \right]  ^{d\left(  L+1\right)  /2-I}}\nonumber\\
&  =\frac{\left\vert \Gamma\left(  I-L\mathcal{D}/2\right)  \right\vert
}{\Gamma\left(  I-Ld/2\right)  }J^{\left(  d\right)  }\lesssim\mathrm{const}%
\times e^{-\frac{\pi L\left\vert \operatorname{Im}\mathcal{D}\right\vert }{4}%
}\left\vert \operatorname{Im}\mathcal{D}\right\vert ^{I-1/2-L\operatorname{Re}%
\mathcal{D}/2}%
\end{align}
These properties of the parametric representation will be used in the examples
in the next Section.

\section{Examples}

In this Section we give several examples of the application of our method to
different loop integrals. Except for the last example, our consideration is
restricted by the one-scale integrals, since several-scale integrals can be
treated by the differential equation method. In all cases, the IBP reduction
has been performed using the \texttt{Mathematica }program, based on Ref.
\cite{Lee2008}. All integrals considered in this paper are known due to some
other techniques. Our representations for one-scale cases, however, differ
from the known ones, being exponentially convergent sums.

\subsection{Three-loop sunrise tadpole: basics of the technique}

Let us consider the three-loop integral%
\begin{equation}
J_{1}^{\left(  \mathcal{D}\right)  }=%
\raisebox{-0.4218cm}{\includegraphics[
trim=0.000000in 0.000000in -0.002646in 0.019570in,
height=1.0456cm,
width=1.1225cm
]%
{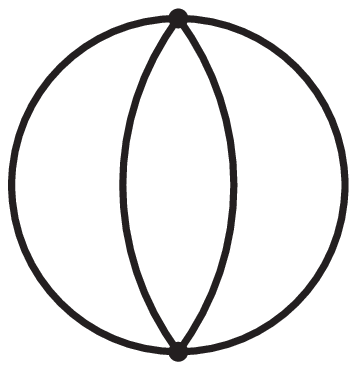}%
}%
=\int\frac{d^{\mathcal{D}}k\,d^{\mathcal{D}}ld^{\mathcal{D}}r}{\pi
^{3\mathcal{D}/2}\left[  k^{2}+1\right]  \left[  l^{2}+1\right]  \left[
r^{2}+1\right]  \left[  \left(  k+l+r\right)  ^{2}+1\right]  }%
\end{equation}
This integral has been investigated in Refs.\cite{Broadhu1992,Broadhurst1996} and the result
is expressed in terms of $_{3}F_{2}\left[  \ldots|1\right]  $. We will follow the path formulated in
the end of Section II.

\begin{enumerate}
\item There is one master integral in subtopologies:%
\begin{equation}
J_{1a}^{\left(  \mathcal{D}\right)  }=%
\raisebox{-0.391cm}{\includegraphics[
trim=0.000000in 0.000000in -0.019448in -0.013839in,
height=0.9951cm,
width=1.0719cm
]%
{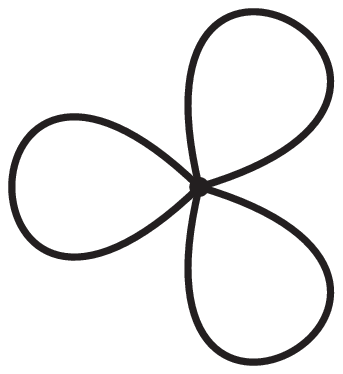}%
}%
=\Gamma^{3}\left(  1-\mathcal{D}/2\right)
\end{equation}

\item The integral $J_{1}^{\left(  \mathcal{D}\right)  }$ is a holomorphic
function in the stripe $S=\left\{  \mathcal{D}|\quad\operatorname{Re}%
\mathcal{D\in}\left[  -2,0\right)  \right\}  $ which can be easily understood
from the parametric representation:%

\begin{equation}
J_{1}^{\left(  \mathcal{D}\right)  }=\Gamma\left(  4-3\mathcal{D}/2\right)
\int\frac{dx_{1}dx_{2}dx_{3}dx_{4}\delta\left(  1-\sum x_{i}\right)  }{\left[
x_{1}x_{2}x_{3}+x_{1}x_{2}x_{4}+x_{1}x_{3}x_{4}+x_{2}x_{3}x_{4}\right]
^{\mathcal{D}/2}} \label{eq:parametric1}%
\end{equation}
It easy to realise that any Euclidean integral with all internal propagators
massive is holomorphic in the whole half-plane $\operatorname{Re}%
\mathcal{D}<0$.

\item The dimensional recurrence reads%
\begin{equation}
J_{1}^{\left(  \mathcal{D}-2\right)  }=-\frac{\left(  3\mathcal{D}-10\right)
_{3}\left(  \mathcal{D}-2\right)  }{128(\mathcal{D}-4)}J_{1}^{\left(
\mathcal{D}\right)  }-\frac{(11\mathcal{D}-38)(\mathcal{D}-2)^{3}%
}{64(\mathcal{D}-4)}J_{1a}^{\left(  \mathcal{D}\right)  }%
\end{equation}
where $\left(  \alpha\right)  _{n}=\alpha\left(  \alpha+1\right)
\ldots\left(  \alpha+n-1\right)  $ denotes the Pochhammer symbol.

\item We choose the summing factor as%
\begin{equation}
\Sigma^{-1}\left(  \mathcal{D}\right)  =4^{\mathcal{D}}\frac{\Gamma\left(
3/2-\mathcal{D}/2\right)  \Gamma\left(  3-3\mathcal{D}/2\right)  }%
{\Gamma\left(  2-\mathcal{D}/2\right)  }%
\end{equation}
Note that the chosen summing factor has neither poles nor zeros in $S$.
Representing $J_{1}^{\left(  \mathcal{D}\right)  }=\Sigma^{-1}\left(
\mathcal{D}\right)  g\left(  \mathcal{D}\right)  $, we obtain%
\begin{align}
g\left(  \mathcal{D}-2\right)   &  =g\left(  \mathcal{D}\right)  +r\left(
\mathcal{D}\right)  \nonumber\\
r\left(  \mathcal{D}\right)   &  =-\frac{\left(  11\mathcal{D}-38\right)
\Gamma^{4}\left(  2-\mathcal{D}/2\right)  }{4^{\mathcal{D}}\Gamma\left(
5/2-\mathcal{D}/2\right)  \Gamma\left(  6-3\mathcal{D}/2\right)  }%
\end{align}
The nonhomogeneous part $r\left(  \mathcal{D}-2k\right)  $ falls down as
$\left(  16/27\right)  ^{k}$ when $k$ goes to $\infty$, so we can write the
solution in the form%
\begin{align}
J_{1}^{\left(  \mathcal{D}\right)  } &  =\omega(z)/\Sigma\left(
\mathcal{D}\right)  +t\left(  \mathcal{D}\right)  =4^{\mathcal{D}}\frac
{\Gamma\left(  3/2-\mathcal{D}/2,3-3\mathcal{D}/2\right)  }{\Gamma\left(
2-\mathcal{D}/2\right)  }\omega(z)\nonumber\\
&  +\frac{1}{16\Gamma\left(  2-\mathcal{D}/2\right)  }\sum_{k=1}^{\infty}%
\frac{\left(  11\mathcal{D}-16-22k\right)  \Gamma^{4}\left(  1+k-\mathcal{D}%
/2\right)  }{\left(  3/2-\mathcal{D}/2\right)  _{k}\left(  3-3\mathcal{D}%
/2\right)  _{3k}}16^{k}\label{eq:general1}%
\end{align}

\item Let us consider analytical properties of functions entering
Eq.(\ref{eq:general1}) in $S$. The integral $J_{1}^{(\mathcal{D})}$ is a
holomorphic function in $S$. When $\operatorname{Im}\mathcal{D\rightarrow
\pm\infty}$, it falls down at least as
\begin{equation}
\left\vert J_{1}^{(\mathcal{D})}\right\vert \lesssim\left\vert \Gamma\left(
4-3\mathcal{D}/2\right)  \right\vert \sim e^{-\frac{3\pi\left\vert
\operatorname{Im}\mathcal{D}\right\vert }{4}}\left\vert \operatorname{Im}%
\mathcal{D}\right\vert ^{7/2-3\operatorname{Re}\mathcal{D}/2}\label{eq:int1}%
\end{equation}
which can be clearly seen from the parametric representation
(\ref{eq:parametric1}). The summing factor is also holomorphic function in $S$
and
\begin{equation}
\left\vert \Sigma\left(  \mathcal{D}\right)  \right\vert \lesssim
e^{\frac{3\pi\left\vert \operatorname{Im}\mathcal{D}\right\vert }{4}%
}\left\vert \operatorname{Im}\mathcal{D}\right\vert ^{-2+3\operatorname{Re}%
\mathcal{D}/2}\label{eq:if1}%
\end{equation}
in the limit $\operatorname{Im}\mathcal{D\rightarrow\pm\infty}$. Finally, the
specific solution $t\left(  \mathcal{D}\right)  $ of the nonhomogeneous
equation is also a holomorphic function in $S$. and
\begin{equation}
\left\vert t\left(  \mathcal{D}\right)  \right\vert \lesssim e^{-\frac
{3\pi\left\vert \operatorname{Im}\mathcal{D}\right\vert }{4}}\left\vert
\operatorname{Im}\mathcal{D}\right\vert ^{\sigma}\label{eq:t1}%
\end{equation}
Note that the limits $\operatorname{Im}\mathcal{D\rightarrow}\pm\infty$
correspond to the limits $z\rightarrow0,\infty$. So, from Eqs.
(\ref{eq:general1}), (\ref{eq:int1}), (\ref{eq:if1}), and (\ref{eq:t1}) we
conclude that
\begin{equation}
\omega(z)\overset{z\rightarrow0,\infty}{\sim}\left\vert \operatorname{Im}%
\mathcal{D}\right\vert ^{\nu},
\end{equation}
where $\nu$ is some real number not essential for our consideration. Taking
into account that
\begin{equation}
\lim\limits_{z\rightarrow\infty}\frac{\left\vert \operatorname{Im}%
\mathcal{D}\right\vert ^{\nu}}{\left\vert z\right\vert ^{\alpha}}%
=\lim\limits_{z\rightarrow0}\frac{\left\vert \operatorname{Im}\mathcal{D}%
\right\vert ^{\nu}}{\left\vert z\right\vert ^{-\alpha}}=0
\end{equation}
for any $\nu$ and any $\alpha>0$, we conclude that $\omega\left(  z\right)  $
is holomorphic function in the extended complex plane of $z$, except, may be
$z=0$ and $z=\infty$ and growing slower than any positive(negative) power of
$\left\vert z\right\vert $ when $z$ tends to infinity (zero). This is
sufficient to claim that $\omega\left(  z\right)  $ is holomorphic function in
the extended complex plane of $z$, being therefore a constant.

\item We can fix this constant by the condition $J_{1}^{\left(  \mathcal{D}%
=0\right)  }=1$, which, e.g., follows from the parametric representation. We
finally obtain%
\begin{align}
J_{1}^{\left(  \mathcal{D}\right)  } &  =4^{\mathcal{D}}\frac{\Gamma\left(
3/2-\mathcal{D}/2,3-3\mathcal{D}/2\right)  }{\Gamma\left(  2-\mathcal{D}%
/2\right)  }\frac{1}{\sqrt{\pi}}\sum_{k=0}^{\infty}\frac{\left(
1+11k/8\right)  \left(  k!\right)  ^{4}}{\left(  3/2\right)  _{k}\left(
3\right)  _{3k}}16^{k}\nonumber\\
&  +\frac{1}{16\Gamma\left(  2-\mathcal{D}/2\right)  }\sum_{k=1}^{\infty}%
\frac{\left(  11\mathcal{D}-16-22k\right)  \Gamma^{4}\left[  k+1-\mathcal{D}%
/2\right]  }{\left(  3/2-\mathcal{D}/2\right)  _{k}\left(  3-3\mathcal{D}%
/2\right)  _{3k}}16^{k}\label{eq:J2inD}%
\end{align}
It is easy to understand that general terms of both sums in Eq.
(\ref{eq:J2inD}) fall down as $\left(  \frac{16}{27}\right)  ^{k}$ thus
providing fast convergence of sums which can be conveniently determined as
$1/\log_{10}\left(  27/16\right)  \approx4.4$ terms per decimal digit. We may
have expressed our result (\ref{eq:J2inD}) in terms of $_{p}F_{q}\left[
\ldots|\frac{16}{27}\right]  $, but we consider the sums to be more treatable
than these functions. In particular, in order to obtain the $\epsilon
$-expansion of $J_{1}^{\left(  4-2\epsilon\right)  }$, we can expand under the
sum sign. It is easy to check that, up to a $O\left(  \epsilon^{0}\right)  $,
this expansion is given by the first term of the second sum:
\begin{equation}
J_{1}^{\left(  4-2\epsilon\right)  }=-\frac{4}{3}\frac{11\epsilon-3}{\left(
2\epsilon-1\right)  \left(  3\epsilon-1\right)  \left(  3\epsilon-2\right)
\left(  \epsilon-1\right)  }\Gamma^{3}\left[  \epsilon\right]  +O\left(
\epsilon\right)
\end{equation}
In fact, the next 5 terms of $\epsilon$-expansion of $J_{1}^{\left(
4-2\epsilon\right)  }$ can be found in Refs.\cite{Broadhu1992,Broadhurst1996}.
In order to demonstrate the fast convergence of the obtained representation we
present here the 40-digit values of coefficients of $J_{1}^{\left(
4-2\epsilon\right)  }$ expansion up to $O\left(  \epsilon^{6}\right)  $
obtained in less than 2 minutes on \texttt{Mathematica}:%
\begin{align}
J_{1}^{\left(  4-2\epsilon\right)  } &
=+2.00000000000000000000000000000000000000\times\epsilon^{-3}\nonumber\\
&  +4.203372677257469503027594126172252080413\times\epsilon^{-2}\nonumber\\
&  +12.15744322207889158433285265524001769362\times\epsilon^{-1}\nonumber\\
&  +10.34393554171240616956324844514078698433\times\epsilon^{0}\nonumber\\
&  +22.40040337934740516273278836060402645950\times\epsilon\nonumber\\
&  -192.9905185401601934396793596956955193083\times\epsilon^{2}\nonumber\\
&  -298.2989210535372236865167423656517391923\times\epsilon^{3}\nonumber\\
&  -3327.931223253862483955917547688685145911\times\epsilon^{4}\nonumber\\
&  -5027.397388051400645438297589076204952450\times\epsilon^{5}\nonumber\\
&  -37321.16865675250679268222655965209414217\times\epsilon^{6}\nonumber\\
&  +O\left(  \epsilon^{7}\right)
\end{align}

\end{enumerate}

\subsection{Four-loop tadpole: Dealing with masless internal lines.}

Let us now consider the following four-loop tadpole integral:%

\begin{equation}
J_{2}^{\left(  \mathcal{D}\right)  }=%
\raisebox{-0.4108cm}{\includegraphics[
trim=0.000000in 0.000000in 0.011219in 0.002844in,
height=1.0214cm,
width=1.0961cm
]%
{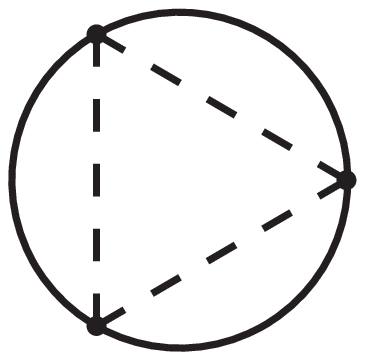}%
}
=\int\frac{d^{\mathcal{D}}k\,d^{\mathcal{D}}l\,d^{\mathcal{D}}%
r\,d^{\mathcal{D}}p}{\pi^{2\mathcal{D}}k^{2}l^{2}r^{2}\left[  \left(
k+p\right)  ^{2}+1\right]  \left[  \left(  l+p\right)  ^{2}+1\right]  \left[
\left(  r+p\right)  ^{2}+1\right]  }%
\end{equation}
The explicit expression for this integral and expansions near $\mathcal{D}=4$
and $\mathcal{D}=3$ can be found in Ref. \cite{BejdSch2006}. In this example
we will demonstrate how to deal with the massless propagators.

\begin{enumerate}
\item There is one master integral in subtopologies:%
\begin{equation}
J_{2a}^{\left(  \mathcal{D}\right)  }=%
\raisebox{-0.2504cm}{\includegraphics[
trim=0.000000in 0.000000in -0.013745in -0.008341in,
height=0.7139cm,
width=1.4014cm
]%
{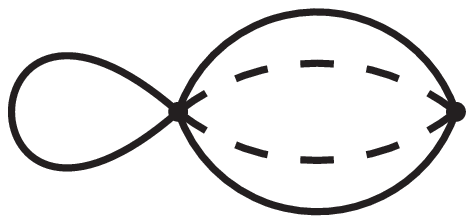}%
}
=-\frac{\Gamma\left(  4-3\mathcal{D}/2\right)  \Gamma^{2}\left(
3-\mathcal{D}\right)  \Gamma^{2}\left(  1-\mathcal{D}/2\right)  \Gamma\left(
\mathcal{D}/2-1\right)  }{\Gamma\left(  6-2\mathcal{D}\right)  }%
\end{equation}

\item Due to the presence of massless propagators, there is an infrared
divergence at $\mathcal{D}=2$ in $J_{2}^{\left(  \mathcal{D}\right)  }$. It is
also easy to see that there is an ultraviolet divergence at $\mathcal{D}=3$.
Thus, the integral is a holomorphic function only in the stripe $S=\left\{
\mathcal{D}|\quad\operatorname{Re}\mathcal{D\in}\left(  2,3\right)  \right\}
$. This is not sufficient for our purposes and we instead consider another
integral, with massive propagators squared:%
\begin{equation}
\tilde{J}_{2}^{\left(  \mathcal{D}\right)  }=%
\raisebox{-0.4108cm}{\includegraphics[
trim=0.000000in 0.000000in 0.011159in 0.002844in,
height=1.0214cm,
width=1.1225cm
]%
{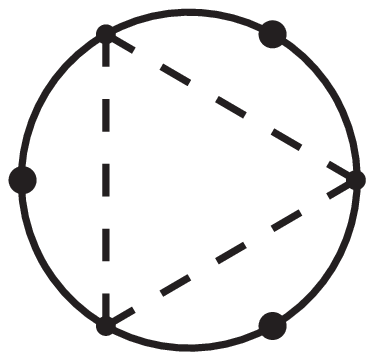}%
}
=\int\frac{d^{\mathcal{D}}k\,d^{\mathcal{D}}l\,d^{\mathcal{D}}%
r\,d^{\mathcal{D}}p}{\pi^{2\mathcal{D}}k^{2}l^{2}r^{2}\left[  \left(
k+p\right)  ^{2}+1\right]  ^{2}\left[  \left(  l+p\right)  ^{2}+1\right]
^{2}\left[  \left(  r+p\right)  ^{2}+1\right]  ^{2}}%
\end{equation}
Due to IBP identities, the original integral $J_{2}^{\left(  \mathcal{D}%
\right)  }$ can be expressed via $\tilde{J}_{2}^{\left(  \mathcal{D}\right)
}$ and $J_{2a}^{\left(  \mathcal{D}\right)  }$ as follows:%
\begin{align}
J_{2}^{\left(  \mathcal{D}\right)  } &  =-\frac{3(3\mathcal{D}%
-11)(3\mathcal{D}-10)}{4(\mathcal{D}-4)(\mathcal{D}-3)^{3}(2\mathcal{D}%
-7)}\tilde{J}_{2}^{\left(  \mathcal{D}\right)  }\nonumber\\
&  -\frac{3(\mathcal{D}-2)(3\mathcal{D}-8)\left(  13\mathcal{D}^{2}%
-88\mathcal{D}+148\right)  }{128(\mathcal{D}-3)^{2}(2\mathcal{D}-7)^{2}}%
J_{2a}^{\left(  \mathcal{D}\right)  }\label{eq:J3viaJ3t}%
\end{align}
Thus, we can restrict ourselves to the calculation of the integral $\tilde
{J}_{2}^{\left(  \mathcal{D}\right)  }$.This integral is obviously a
holomorphic function of $\mathcal{D}$ when $\operatorname{Re}\mathcal{D\in
}\left(  2,4\frac{1}{2}\right)  $. This is sufficient for our purposes. So we
choose the basic stripe as $S=\left\{  \mathcal{D}|\quad\operatorname{Re}%
\mathcal{D\in}\left(  2,4\right]  \right\}  $. Let us make one observation
which essentially simplifies the following consideration. The two terms in the
right-hand side of Eq. (\ref{eq:J3viaJ3t}) have poles of the third order at
$\mathcal{D}=3$, while the integral $J_{2}^{\left(  \mathcal{D}\right)  }$
clearly has only first-order pole at $\mathcal{D}=3$. Thus, from Eq.
(\ref{eq:J3viaJ3t}), we can extract two terms of expansion of $\tilde{J}%
_{2}^{\left(  \mathcal{D}\right)  }$ near $\mathcal{D}=3$:%
\begin{equation}
\tilde{J}_{2}^{\left(  3-2\epsilon\right)  }=\frac{\pi^{2}}{4}+\epsilon
\frac{\pi^{2}}{4}\left(  11-4\gamma-8\ln2\right)  +O\left(  \epsilon\right)
\label{eq:J3in3d}%
\end{equation}

\item The dimensional recurrence for $\tilde{J}_{2}^{\left(  \mathcal{D}%
\right)  }$ reads%
\begin{align}
\tilde{J}_{2}^{\left(  \mathcal{D}\right)  }  &  =-\frac{4(2\mathcal{D}%
-8)_{4}(\mathcal{D}-3)^{3}(\mathcal{D}-1)_{2}}{3(3\mathcal{D}-11)_{5}}%
\tilde{J}_{2}^{\left(  \mathcal{D}+2\right)  }+R\left(  \mathcal{D}\right)
\nonumber\\
R\left(  \mathcal{D}\right)   &  =-\frac{\left(  3\mathcal{D}/2-3\right)
_{4}(\mathcal{D}-4)_{5}(\mathcal{D}-3)^{2}}{96(3\mathcal{D}-11)_{5}%
(2\mathcal{D}-7)_{5}}(3299\mathcal{D}^{6}-59493\mathcal{D}^{5}%
+444098\mathcal{D}^{4}\nonumber\\
&  -1756164\mathcal{D}^{3}+3879800\mathcal{D}^{2}-4540224\mathcal{D}%
+2198784)J_{2a}^{\left(  \mathcal{D}+2\right)  }%
\end{align}

\item It is easy to establish from the explicit form of known integral
$J_{2a}^{\left(  \mathcal{D}\right)  }$ that in this case $R_{-}\left(
\mathcal{D}\right)  =0$. The summing factor obeys the equation
\begin{equation}
\frac{\Sigma\left(  \mathcal{D}+2\right)  }{\Sigma\left(  \mathcal{D}\right)
}=-\frac{4(2\mathcal{D}-8)_{4}(\mathcal{D}-3)^{3}(\mathcal{D}-1)_{2}%
}{3(3\mathcal{D}-11)_{5}}%
\end{equation}
As we noticed above, we can use different forms of the summing factor and we
choose%
\begin{align}
\Sigma\left(  \mathcal{D}\right)   &  =\frac{\sin\left(  \frac{\pi}%
{2}\mathcal{D}-\frac{2\pi}{3}\right)  \sin\left(  \frac{\pi}{2}\mathcal{D}%
-\frac{\pi}{3}\right)  \sin^{2}\left(  \frac{\pi\mathcal{D}}{2}\right)
}{2^{\mathcal{D}}\sin\left(  \frac{\pi}{2}\mathcal{D}-\frac{5\pi}{6}\right)
\sin\left(  \frac{\pi}{2}\mathcal{D}-\frac{\pi}{6}\right)  }\label{eq:I3}\\
&  \times\frac{\Gamma\left(  7-\frac{3\mathcal{D}}{2}\right)  \Gamma
^{2}\left(  \frac{\mathcal{D}}{2}-\frac{3}{2}\right)  \Gamma(\mathcal{D}%
-1)}{\Gamma(9-2\mathcal{D})\Gamma(5-\mathcal{D})\Gamma\left(  \frac
{3\mathcal{D}}{2}-\frac{11}{2}\right)  }.\nonumber
\end{align}
Our choice of the summing factor needs some explanations. When choosing the
summing factor, we used the following "rules of thumb":

\begin{enumerate}
\item \label{str:a}"No undetermined singularities in basic stripe": The
function $\Sigma\left(  \mathcal{D}\right)  $ is chosen in such a way that
$\Sigma\left(  \mathcal{D}\right)  J_{2}^{\left(  \mathcal{D}\right)  }$ has
only known singularities in $S$. We did not eliminate the simple pole at
$\mathcal{D}=3$ because the singular part of $\Sigma\left(  \mathcal{D}%
\right)  J_{2}^{\left(  \mathcal{D}\right)  }$ near $\mathcal{D}=3$ is known
due to Eq.(\ref{eq:J3in3d}).

\item "Good behaviour at infinity": We require that the function
$\Sigma\left(  \mathcal{D}\right)  J_{2}^{\left(  \mathcal{D}\right)  }$ obey
the following constraint%
\begin{equation}
\frac{\Sigma\left(  \mathcal{D}\right)  J_{2}^{\left(  \mathcal{D}\right)  }%
}{\exp\left[  \pi\left\vert \mathcal{D}\right\vert \right]  }\overset
{\operatorname{Im}\mathcal{D}\rightarrow\pm\infty}{\mathcal{\longrightarrow}%
}0\label{eq:constr1_3}%
\end{equation}
when $\operatorname{Re}\,\mathcal{D}\in\left(  2,4\right]  $. It is very easy
to establish from the Feynman parametrization, that%
\[
J_{3}^{\left(  \mathcal{D}\right)  }\overset{\operatorname{Im}\mathcal{D}%
\rightarrow\pm\infty}{\lesssim}\Gamma\left[  9-2\mathcal{D}\right]  \left\vert
\operatorname{Im}\mathcal{D}\right\vert ^{\nu},
\]
so we can easily obtain the desired estimate for $\Sigma\left(  \mathcal{D}%
\right)  J_{2}^{\left(  \mathcal{D}\right)  }$. We also require $r_{+}\left(
\mathcal{D}\right)  =\Sigma\left(  \mathcal{D}\right)  R\left(  \mathcal{D}%
\right)  $ to obey the same constraint%
\begin{equation}
\frac{r_{+}\left(  \mathcal{D}\right)  }{\exp\left[  \pi\left\vert
\mathcal{D}\right\vert \right]  }\overset{\operatorname{Im}\mathcal{D}%
\rightarrow\pm\infty}{\mathcal{\longrightarrow}}0\label{eq:constr2_3}%
\end{equation}
for $\operatorname{Re}\,\mathcal{D}>2$.

\item "Pole minimization": The function $r_{+}\left(  \mathcal{D}\right)
=\Sigma\left(  \mathcal{D}\right)  R\left(  \mathcal{D}\right)  $ has some
poles. We try to minimize their orders and/or number in the region
$\operatorname{Re}\,\mathcal{D}>2$ by adding some factors of the form
$\sin\left(  \left(  \mathcal{D}-d\right)  \pi/2\right)  $. This requirement
explains appearance of sin functions in the numerator in Eq. (\ref{eq:I3}). At
some point we were not able to add sin factors due to the condition b.
\end{enumerate}

The pole structure of
\begin{align}
r_{+}\left(  \mathcal{D}\right)   &  =-\frac{3\pi^{5/2}2^{2\mathcal{D}%
-15}\mathcal{D}\cos\left(  \frac{3\pi\mathcal{D}}{2}\right)  \Gamma
(7-3\mathcal{D})\Gamma\left(  4-\frac{3\mathcal{D}}{2}\right)  \Gamma\left(
\frac{\mathcal{D}}{2}-\frac{1}{2}\right)  ^{2}}{\sin(\pi\mathcal{D}%
)\tan\left(  \frac{\pi\mathcal{D}}{2}-\frac{\pi}{6}\right)  \tan\left(
\frac{\pi\mathcal{D}}{2}+\frac{\pi}{6}\right)  \Gamma(8-2\mathcal{D}%
)^{2}\Gamma\left(  \frac{\mathcal{D}}{2}+1\right)  }\nonumber\\
&  \times(3299\mathcal{D}^{6}-59493\mathcal{D}^{5}+444098\mathcal{D}%
^{4}-1756164\mathcal{D}^{3}\nonumber\\
&  +3879800\mathcal{D}^{2}-4540224\mathcal{D}+2198784)\label{eq:r3}%
\end{align}
is the following. There are 5 series of simple poles in%
\begin{gather}
\mathcal{D}=\mathcal{D}_{1}\left(  k\right)  ,\,\mathcal{D}_{2}\left(
k\right)  ,\,\mathcal{D}_{3}\left(  k\right)  ,\,\mathcal{D}_{4}\left(
k\right)  ,\mathcal{D}_{5}\left(  k\right)  ,\qquad k=0,1,\ldots\nonumber\\
\mathcal{D}_{1}\left(  k\right)  =2\tfrac{1}{3}+2k,\quad\mathcal{D}_{2}\left(
k\right)  =2\tfrac{2}{3}+2k,\nonumber\\
\mathcal{D}_{3}\left(  k\right)  =3\tfrac{1}{3}+2k,\quad\mathcal{D}_{4}\left(
k\right)  =3\tfrac{2}{3}+2k,\nonumber\\
\mathcal{D}_{5}\left(  k\right)  =4+2k,\label{eq:polestructure3}%
\end{gather}
and also a single pole at $\mathcal{D}=3$. This pole structure is depicted in
Fig.\ref{fig:tadpole4_r_poles}.%
\begin{figure}
[ptb]
\begin{center}
\includegraphics[
height=2.4344in,
width=4.062in
]%
{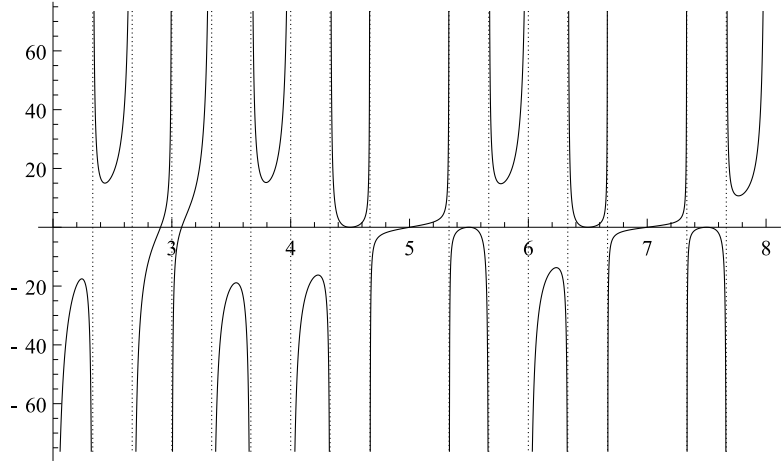}%
\caption{The function $r_{+}\left(  \mathcal{D}\right)  $.}%
\label{fig:tadpole4_r_poles}%
\end{center}
\end{figure}
The general solution of the recurrence relation reads%
\begin{equation}
\tilde{J}_{2}^{\left(  \mathcal{D}\right)  }=\Sigma^{-1}\left(  \mathcal{D}%
\right)  \left[  \omega\left(  z\right)  +\sum_{k=0}^{\infty}r_{+}\left(
\mathcal{D}+2k\right)  \right]  ,\label{eq:general3}%
\end{equation}
where $\omega\left(  z\right)  =\omega\left(  e^{i\pi\mathcal{D}}\right)  $ is
the arbitrary periodic function of $\mathcal{D}$ we need to fix.

\item Let us now analyse the singularities of $\omega\left(  z\right)  $.
Rewriting Eq.(\ref{eq:general3}) as%
\begin{equation}
\omega\left(  z\right)  =\Sigma\left(  \mathcal{D}\right)  \tilde{J}%
_{2}^{\left(  \mathcal{D}\right)  }-\sum_{k=0}^{\infty}r_{+}\left(
\mathcal{D}+2k\right)  ,\label{eq:om3}%
\end{equation}
we see that singularities of this function are determined by the singularities
of the terms in the right-hand side. Due to our choice of the summing factor
(see \ref{str:a}), the first term has only known singularities in $S$. Namely,
it has simple pole at $\mathcal{D}=3$ with the residue, determined from Eq.
(\ref{eq:J3in3d}):%
\begin{equation}
\Sigma\left(  3-2\epsilon\right)  \tilde{J}_{2}^{\left(  3-2\epsilon\right)
}=\frac{3\pi^{5/2}}{256\epsilon}+O\left(  \epsilon^{0}\right)
\end{equation}
From the pole structure of $r_{+}\left(  \mathcal{D}\right)  $ we conclude that the first term of
the sum in the right-hand side of Eq.(\ref{eq:om3}) also has a pole at $\mathcal{D}=3$,
while the other terms of the sum have no
singularities in this point. Moreover,%
\begin{equation}
r_{+}\left(  3-2\epsilon\right)  =\frac{3\pi^{5/2}}{256\epsilon}+O\left(
\epsilon^{0}\right)  ,
\end{equation}
so the singularities of the right-hand side of Eq.(\ref{eq:om3}) at
$\mathcal{D}=3$ cancel. Thus, $\omega\left(  z\right)  $ has no singularities
in $z=\exp\left[  i\pi\times3\right]  =-1$. The five series of poles
(\ref{eq:polestructure3}) give rise to the poles of $\omega\left(  z\right)  $
in%
\begin{equation}
z=e^{i\pi/3},e^{i2\pi/3},e^{-i2\pi/3},e^{-i\pi/3},1.
\end{equation}
From the constraints (\ref{eq:constr1_3}) and (\ref{eq:constr2_3}) we conclude
that $\omega\left(  z\right)  $ is regular at $z=0,\infty$. Thus, the general
form of $\omega\left(  z\right)  $, mimicking the structure of the poles in
the right-hand side of Eq. (\ref{eq:om3}), is the following%
\begin{equation}
\omega\left(  z\right)  =a_{0}+\frac{a_{1}}{z-e^{i\pi/3}}+\frac{a_{2}%
}{z-e^{i2\pi/3}}+\frac{a_{3}}{z-e^{-i2\pi/3}}+\frac{a_{4}}{z-e^{-i\pi/3}%
}+\frac{a_{5}}{z-1}%
\end{equation}
It is convenient to rewrite this form in terms of cotangents. Using the
identity%
\begin{equation}
\cot\left(  \frac{\pi}{2}\left(  \mathcal{D}-\mathcal{D}_{0}\right)  \right)
=i\frac{z+e^{i\pi\mathcal{D}_{0}}}{z-e^{i\pi\mathcal{D}_{0}}}%
\end{equation}
we represent%
\begin{align}
\omega\left(  z\right)   &  =b+b_{1}\cot\tfrac{\pi}{2}\left(  \mathcal{D}%
-2\tfrac{1}{3}\right)  +b_{2}\cot\tfrac{\pi}{2}\left(  \mathcal{D}-2\tfrac
{2}{3}\right)  \nonumber\\
&  +b_{3}\cot\tfrac{\pi}{2}\left(  \mathcal{D}-3\tfrac{1}{3}\right)
+b_{4}\cot\tfrac{\pi}{2}\left(  \mathcal{D}-3\tfrac{2}{3}\right)  +b_{5}%
\cot\tfrac{\pi}{2}\left(  \mathcal{D}-4\right)
\end{align}
The constants $b_{1-5}$ are fixed by the singularities of the sum in the
right-hand side of Eq.(\ref{eq:om3}):%
\begin{equation}
b_{i}=-\frac{\pi}{2}\sum_{k=0}^{\infty}~\operatorname*{Res}%
\limits_{\mathcal{D}=\mathcal{D}_{i}\left(  k\right)  }r_{+}\left(
\mathcal{D}\right)  ,
\end{equation}
where $\mathcal{D}_{i}\left(  k\right)  $ are determined in
Eq.(\ref{eq:polestructure3}). They can be expressed as linear combinations of
$_{p}F_{q}\left[  \ldots\left\vert \frac{2^{14}}{3^{9}}\right.  \right]  $.
Using a guess $b_{i}=f_{i}\pi^{7/2}$ with $f_{i}$ being rationals, we were
able to establish the simple form of these coefficients:%
\begin{equation}
b_{1}\overset{\mathrm{N}}{=}-b_{2}\overset{\mathrm{N}}{=}-b_{3}\overset
{\mathrm{N}}{=}b_{4}\overset{\mathrm{N}}{=}-\tfrac{1}{4}b_{5}\overset
{\mathrm{N}}{=}-\frac{3\pi^{7/2}}{32}.
\end{equation}
where $\overset{\mathrm{N}}{=}$ denotes the equality checked numerically with
at least $10^{3}$ digits.

\item The last coefficient $b$ in $\omega\left(  z\right)  $ can be shown to be equal to zero
    by cosidering the limit of the right-hand side of Eq.(\ref{eq:om3}) at
    $\mathcal{D}\rightarrow3$ and using Eq. (\ref{eq:J3in3d}). Note that in this limit only the
    first term in the sum contribute. Thus, we
obtain finally%
\begin{align}
\tilde{J}_{2}^{\left(  \mathcal{D}\right)  } &  =\Sigma^{-1}\left(
\mathcal{D}\right)  \left[  -\frac{3\pi^{7/2}}{32}\left(  \cot\tfrac{\pi}%
{2}\left(  \mathcal{D}-2\tfrac{1}{3}\right)  -\cot\tfrac{\pi}{2}\left(
\mathcal{D}-2\tfrac{2}{3}\right)  \right.  \right.  \nonumber\\
&  \left.  -\cot\tfrac{\pi}{2}\left(  \mathcal{D}-3\tfrac{1}{3}\right)
+\cot\tfrac{\pi}{2}\left(  \mathcal{D}-3\tfrac{2}{3}\right)  -4\cot\tfrac{\pi
}{2}\left(  \mathcal{D}-4\right)  \right)  \nonumber\\
&  \left.  +\sum_{k=0}^{\infty}r_{+}\left(  \mathcal{D}+2k\right)  \right]
,\label{eq:final2}%
\end{align}
where $\Sigma^{-1}\left(  \mathcal{D}\right)  $ and $r_{+}\left(
\mathcal{D}\right)  $ are determined in Eqs. (\ref{eq:I3}) and (\ref{eq:r3}),
respectively. Again, we can express the sum via $_{p}F_{q}\left[
\ldots\left\vert \frac{2^{14}}{3^{9}}\right.  \right]  $. Thus, the rate of
convergence in representation (\ref{eq:final2}) is $1/\log_{10}\left(
3^{9}/2^{14}\right)  \approx12.5$ terms per decimal digit. We have checked
numerically the agreement of our representation with the known expansions
around $\mathcal{D}=4$ and $\mathcal{D}=3$.
\end{enumerate}

\subsection{Two-loop massless onshell nonplanar vertex: Two-directed
summation}

This example demonstrates how the method works in the case when both $R_{+}$
and $R_{-}$ are not zero. Let us consider the integral%
\begin{equation}
J_{3}^{\left(  \mathcal{D}\right)  }=%
\raisebox{-0.4811cm}{\includegraphics[
trim=0.000000in 0.000000in -0.004791in 0.000000in,
height=1.1642cm,
width=1.2894cm
]%
{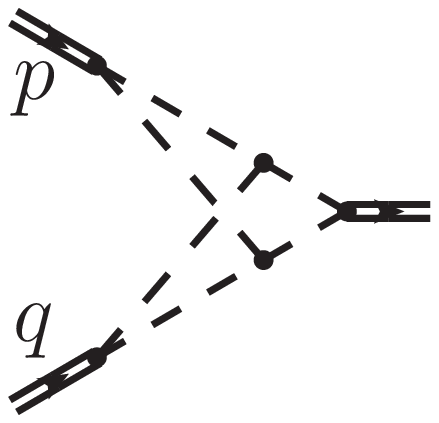}%
}
=\int\frac{\,d^{\mathcal{D}}l\,d^{\mathcal{D}}r}{\pi^{\mathcal{D}}l^{2}%
r^{2}\left(  l-p\right)  ^{2}\left(  r-q\right)  ^{2}\left(  l-r-p\right)
^{2}\left(  r-l-q\right)  ^{2}}%
\end{equation}
with%
\begin{equation}
p^{2}=q^{2}=0,\quad\left(  p+q\right)  ^{2}=-1.
\end{equation}
In contrast to the previous examples, here all scalar products are pseudoeuclidean, e.g.,
$l^{2}=l_{0}^{2}-\mathbf{l}^{2}$. This integral in arbitrary dimension has been calculated in Ref.
\cite{GehHuMa2005} and the result is expressed via $_{q+1}F_{q}\left[  \ldots|1\right]  $ with
$q=3,4$. Let us again apply our technique.

\begin{enumerate}
\item There are two master integrals in the subtopologies:%
\begin{align}
J_{3a}^{\left(  \mathcal{D}\right)  }  &  =%
\raisebox{-0.2285cm}{\includegraphics[
trim=0.000000in 0.000000in -0.002702in 0.019562in,
height=0.6612cm,
width=1.6321cm
]%
{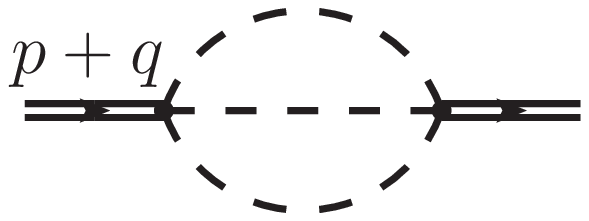}%
}
=\frac{\Gamma(3-\mathcal{D})\Gamma\left(  \frac{\mathcal{D}}{2}-1\right)
^{3}}{\Gamma\left(  \frac{3\mathcal{D}}{2}-3\right)  }\\
J_{3b}^{\left(  \mathcal{D}\right)  }  &  =%
\raisebox{-0.4811cm}{\includegraphics[
trim=0.000000in 0.000000in 0.019547in -0.011014in,
height=1.173cm,
width=1.274cm
]%
{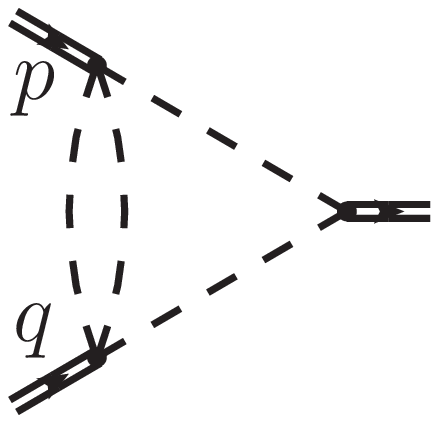}%
}
=-\frac{\Gamma(4-\mathcal{D})\Gamma\left(  2-\frac{\mathcal{D}}{2}\right)
\Gamma\left(  \frac{\mathcal{D}}{2}-1\right)  ^{2}\Gamma(\mathcal{D}-3)^{2}%
}{\Gamma(\mathcal{D}-2)\Gamma\left(  \frac{3\mathcal{D}}{2}-4\right)  }%
\end{align}

\item The integral has collinear divergences at $\mathcal{D}=4$ and
ultraviolet divergence at $\mathcal{D}=6$. So, we choose the basic stripe as
$S=\left\{  \mathcal{D}|\quad\operatorname{Re}\mathcal{D\in}\left(
4,6\right]  \right\}  $. Thorough analysis of Feynman parametrization
justifies that $J_{3}^{\left(  \mathcal{D}\right)  }$ is finite when
$\operatorname{Re}\mathcal{D\in}\left(  4,6\right)  $. We also note that at
$\mathcal{D}=6$ the integral has simple pole.

\item The dimensional recurrence reads%
\begin{align}
J_{3}^{\left(  \mathcal{D}\right)  }  &  =-\frac{(2\mathcal{D}-7)_{4}%
}{(\mathcal{D}-4)_{2}}J_{3}^{\left(  \mathcal{D}+2\right)  }+R_{+}\left(
\mathcal{D}\right)  +R_{-}\left(  \mathcal{D}+2\right) \nonumber\\
R_{+}\left(  \mathcal{D}\right)   &  =-\frac{2\left(  43\mathcal{D}%
^{4}-478\mathcal{D}^{3}+1963\mathcal{D}^{2}-3530\mathcal{D}+2352\right)
}{(\mathcal{D}-3)\left(  \mathcal{D}-4\right)  ^{3}}J_{3a}^{\left(
\mathcal{D}\right)  }\nonumber\\
R_{-}\left(  \mathcal{D}+2\right)   &  =-\frac{2\left(  37\mathcal{D}%
^{3}-313\mathcal{D}^{2}+858\mathcal{D}-752\right)  }{(3\mathcal{D}-8)\left(
\mathcal{D}-4\right)  ^{2}}J_{3b}^{\left(  \mathcal{D}\right)  }.
\end{align}
Here we used the lowering recurrence relation, thus, appearance of
$J_{3a,b}^{\left(  \mathcal{D}\right)  }$ in the right-hand side. It is easy
to check, that $R_{\pm}\left(  \mathcal{D}\right)  $ satisfy the conditions
(\ref{eq:Rpm}).

\item We choose the summing factor as%
\begin{equation}
\Sigma\left(  \mathcal{D}\right)  =4^{\mathcal{D}}(4-\mathcal{D})\sin\left(
\frac{\pi}{2}(\mathcal{D}-5)\right)  \sin^{2}\left(  \frac{\pi}{2}%
(\mathcal{D}-4)\right)  \Gamma\left(  \mathcal{D}-\frac{7}{2}\right)
.\label{eq:integrating4}%
\end{equation}
Note the quadratic zero at $\mathcal{D}=6$, which provides that
\begin{equation}
\lim_{\mathcal{D}\rightarrow6}\Sigma\left(  \mathcal{D}\right)  J_{3}^{\left(
\mathcal{D}\right)  }=0,\label{eq:cond4}%
\end{equation}
and, in particular, that the function $\Sigma\left(  \mathcal{D}\right)
J_{4}^{\left(  \mathcal{D}\right)  }$ is regular at $\mathcal{D}=6$. The
general solution reads%
\begin{align}
J_{3}^{\left(  \mathcal{D}\right)  } &  =\Sigma^{-1}\left(  \mathcal{D}%
\right)  \left[  \omega\left(  \mathcal{D}\right)  +\sum_{k=0}^{\infty}%
r_{+}\left(  \mathcal{D}+2k\right)  -\sum_{k=0}^{\infty}r_{-}\left(
\mathcal{D}-2k\right)  \right]  ,\label{eq:J4_res}\\
r_{+}\left(  \mathcal{D}\right)   &  =\frac{\sqrt{\pi}2^{\mathcal{D}}\sin
(\pi\mathcal{D})\Gamma\left(  \frac{3}{2}-\frac{\mathcal{D}}{2}\right)
\Gamma\left(  \frac{\mathcal{D}}{2}-2\right)  ^{2}\Gamma\left(  \mathcal{D}%
-\frac{7}{2}\right)  }{(\mathcal{D}-3)\Gamma\left(  \frac{3\mathcal{D}}%
{2}-3\right)  }\nonumber\\
&  \times\left(  43\mathcal{D}^{4}-478\mathcal{D}^{3}+1963\mathcal{D}%
^{2}-3530\mathcal{D}+2352\right)  ,\nonumber\\
r_{-}\left(  \mathcal{D}\right)   &  =-\frac{\pi^{2}2^{2\mathcal{D}-6}%
\Gamma\left(  \frac{\mathcal{D}}{2}-3\right)  \Gamma\left(  \mathcal{D}%
-\frac{11}{2}\right)  }{(5-\mathcal{D})\Gamma\left(  \frac{3\mathcal{D}}%
{2}-6\right)  }\left(  37\mathcal{D}^{3}-535\mathcal{D}^{2}+2554\mathcal{D}%
-4016\right)  \nonumber
\end{align}

\item Due to our choice of the summing factor, the function $r_{+}\left(
\mathcal{D}\right)  $ does not have any singularities in the region
$\operatorname{Re}\mathcal{D}>4$. The function $r_{-}\left(  \mathcal{D}%
\right)  $ has simple poles at%
\begin{align}
\mathcal{D} &  =\mathcal{D}_{1}\left(  k\right)  ,\mathcal{D}_{2}\left(
k\right)  ,5,6\nonumber\\
\mathcal{D}_{1}\left(  k\right)   &  =5\tfrac{1}{2}-2k,\quad\mathcal{D}%
_{2}\left(  k\right)  =4\tfrac{1}{2}-2k,\quad k=0,1,\ldots
\end{align}
Taking into account that $\Sigma\left(  \mathcal{D}\right)  J_{4}^{\left(
\mathcal{D}\right)  }$ is holomorphic in $S$, we obtain%
\begin{gather}
\omega\left(  \mathcal{D}\right)  =b+b_{1}\cot\tfrac{\pi}{2}\left(
\mathcal{D}-5\tfrac{1}{2}\right)  +b_{2}\cot\tfrac{\pi}{2}\left(
\mathcal{D}-4\tfrac{1}{2}\right)  \nonumber\\
+b_{3}\cot\tfrac{\pi}{2}\left(  \mathcal{D}-5\right)  +b_{4}\cot\tfrac{\pi}%
{2}\left(  \mathcal{D}-6\right)  ,\nonumber\\
b_{1}=\frac{\pi}{2}\sum_{k=0}^{\infty}\operatorname*{Res}\limits_{\mathcal{D}%
=\mathcal{D}_{1}\left(  k\right)  }r_{-}\left(  \mathcal{D}-2k\right)  ,\quad
b_{2}=\frac{\pi}{2}\sum_{k=0}^{\infty}\operatorname*{Res}\limits_{\mathcal{D}%
=\mathcal{D}_{2}\left(  k\right)  }r_{-}\left(  \mathcal{D}-2k\right)
,\nonumber\\
b_{3}=\frac{\pi}{2}\operatorname*{Res}\limits_{\mathcal{D}=5}r_{-}\left(
\mathcal{D}\right)  =256\pi^{7/2},\quad b_{4}=\frac{\pi}{2}\operatorname*{Res}%
\limits_{\mathcal{D}=6}r_{-}\left(  \mathcal{D}\right)  =1280\pi
^{7/2}\label{eq:om4}%
\end{gather}
Again, making a guess $b_{1,2}=f_{1,2}\pi^{7/2}$, supported by the values of
$b_{3,4}$, we check with the accuracy as high as $10^{3}$digits that%
\begin{equation}
b_{1}\overset{\mathrm{N}}{=}b_{2}\overset{\mathrm{N}}{=}-256\pi^{7/2}%
\label{eq:b1b2_4}%
\end{equation}

\item The constant $b$ can be fixed using the condition (\ref{eq:cond4}) and
the properties of $r_{\pm}$:%
\[
r_{+}\left(  6+2k\right)  =0,\quad r_{-}\left(  6-2\epsilon\right)
-\frac{2560\pi^{5/2}}{-2\epsilon}=O\left(  \epsilon\right)
\]
We obtain%
\begin{equation}
b=\sum_{k=0}^{\infty}r_{-}\left(  4-2k\right)  \overset{\mathrm{N}}%
{=}0.\label{eq:b_4}%
\end{equation}
Therefore, Eqs. (\ref{eq:J4_res}), (\ref{eq:integrating4}), (\ref{eq:om4}), (\ref{eq:b1b2_4}),
and (\ref{eq:b_4}) give us the result for $J_{3}^{\left( \mathcal{D}\right)  }$. Note that the
sums of $r_{\pm}$ in Eq. (\ref{eq:J4_res}) can be expressed via $\,_{5}F_{4}\left[
\ldots|-\frac {16}{27}\right]  $ and $_{4}F_{3}\left[  \ldots|\frac{27}{64}\right]  $, respectively.
The arguments of the hypergeometric functions determine the rate of convergency of the
sums: the first converges with the rate $4.5$ terms per decimal digit, while the second
converges with the rate $2.7$ terms per decimal digit. We have checked numerically the
agreement of our result for the $\epsilon$-expansion of $J_{3}^{\left(  4-2\epsilon\right)  }$
with the corresponding result in Ref. \cite{GehHuMa2005}.
\end{enumerate}

\subsection{Two-loop two-masses tadpole: Two-scale case}

Let us now examine how the method works for the integrals with several scales.
Consider the integral%
\begin{equation}
J_{4}^{\left(  \mathcal{D}\right)  }=%
\raisebox{-0.4327cm}{\includegraphics[
trim=0.000000in 0.000000in -0.011132in 0.008393in,
height=1.0719cm,
width=1.0719cm
]%
{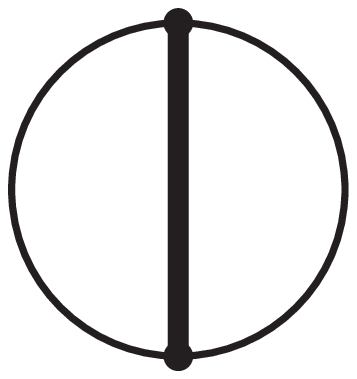}%
}
=\int\frac{\,d^{\mathcal{D}}l\,d^{\mathcal{D}}r}{\pi^{\mathcal{D}}\left[
l^{2}+1\right]  \left[  r^{2}+1\right]  \left[  \left(  l-r\right)  ^{2}%
+m^{2}\right]  }%
\end{equation}
Of course, we can obtain the differential equation with respect to $m$ and put
the initial condition in the point $m=1$. However, we can also proceed as before.

\begin{enumerate}
\item Master integrals in subtopologies are%
\begin{equation}
J_{4a}^{\left(  \mathcal{D}\right)  }=%
\raisebox{-0.1494cm}{\includegraphics[
trim=0.000000in 0.000000in -0.005497in -0.019537in,
height=0.5096cm,
width=1.0719cm
]%
{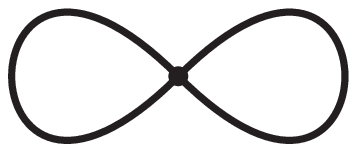}%
}
=m^{\mathcal{D}-4}%
\raisebox{-0.1604cm}{\includegraphics[
trim=0.000000in 0.000000in -0.016621in -0.002767in,
height=0.536cm,
width=1.0961cm
]%
{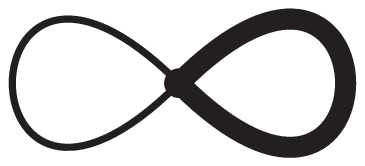}%
}
=\Gamma^{2}\left[  1-\mathcal{D}/2\right]
\end{equation}

\item The basic stripe can be chosen, e.g., as $S=\left\{  \mathcal{D}%
|\quad\operatorname{Re}\mathcal{D\in}\left[  0,2\right)  \right\}  $.

\item The dimensional recurrence reads%
\begin{equation}
J_{4}^{\left(  \mathcal{D}\right)  }=\frac{m^{2}\left(  4-m^{2}\right)
}{\left(  \mathcal{D}-2\right)  \left(  \mathcal{D}-3\right)  }J_{4}^{\left(
\mathcal{D}-2\right)  }+\frac{\left(  m^{2}-2-2m^{\mathcal{D}-2}\right)
}{\left(  \mathcal{D}-2\right)  \left(  \mathcal{D}-3\right)  }\Gamma
^{2}\left(  2-\mathcal{D}/2\right)
\end{equation}

\item Let us assume that $m<2$ for the moment. The result for $m>2$ can be
obtained by the analytical continuation. We choose the summing factor as
\begin{equation}
\Sigma\left(  \mathcal{D}\right)  =\frac{m^{-\mathcal{D}}\left(
4-m^{2}\right)  ^{-\mathcal{D}/2}}{\Gamma(2-\mathcal{D})}%
\end{equation}
After the replacement $J_{4}^{\left(  \mathcal{D}\right)  }=g\left(
\mathcal{D}\right)  /\Sigma\left(  \mathcal{D}\right)  $ we obtain%
\begin{align}
g\left(  \mathcal{D}\right)   &  =g\left(  \mathcal{D}-2\right)  +r\left(
\mathcal{D}\right)  \nonumber\\
r\left(  \mathcal{D}\right)   &  =\frac{\left(  m^{2}-2-2m^{\mathcal{D}%
-2}\right)  \Gamma^{2}\left(  2-\mathcal{D}/2\right)  }{m^{\mathcal{D}}\left(
4-m^{2}\right)  ^{\mathcal{D}/2}\Gamma(4-\mathcal{D})}%
\end{align}
The limit
\begin{equation}
\lim_{\mathcal{D\rightarrow-\infty}}\frac{r\left(  \mathcal{D}+2\right)
}{r\left(  \mathcal{D}\right)  }=\left(  1-\frac{m^{2}}{4}\right)  \max\left(
1,m^{2}\right)  \leqslant1
\end{equation}
shows that $r\left(  \mathcal{D}\right)  =r_{-}\left(  \mathcal{D}\right)  $
and allows us to represent the general solution in the form%
\begin{align*}
J_{4}^{\left(  \mathcal{D}\right)  } &  =\Sigma^{-1}\left(  \mathcal{D}%
\right)  \omega\left(  z\right)  +t\left(  \mathcal{D}\right)  \\
t\left(  \mathcal{D}\right)   &  =\frac{\Gamma\left(  1-\mathcal{D}/2\right)
\Gamma\left(  2-\mathcal{D}/2\right)  }{2\left(  3-\mathcal{D}\right)  }\\
&  \times\sum_{k=0}^{\infty}\left(  1-\frac{m^{2}}{4}\right)  ^{k}\left(
\left(  m^{2}-2\right)  m^{2k}-2m^{\mathcal{D}-2}\right)  \frac{\left(
2-\mathcal{D}/2\right)  _{k}}{\left(  5/2-\mathcal{D}/2\right)  _{k}}\\
&  =\frac{\Gamma\left(  1-\mathcal{D}/2\right)  \Gamma\left(  2-\mathcal{D}%
/2\right)  }{2\left(  3-\mathcal{D}\right)  }\left(  \left(  m^{2}-2\right)
\,_{2}F_{1}\left[
\genfrac{}{}{0pt}{1}{1,2-\mathcal{D}/2}{5/2-\mathcal{D}/2}%
\left\vert m^{2}-\frac{m^{4}}{4}\right.  \right]  \right.  \\
&  \left.  -2m^{\mathcal{D}-2}\,_{2}F_{1}\left[
\genfrac{}{}{0pt}{1}{1,2-\mathcal{D}/2}{5/2-\mathcal{D}/2}%
\left\vert 1-\frac{m^{2}}{4}\right.  \right]  \right)
\end{align*}

\item The functions $\Sigma\left(  \mathcal{D}\right)  J_{4}^{\left(
\mathcal{D}\right)  }$ and $\Sigma\left(  \mathcal{D}\right)  t\left(
\mathcal{D}\right)  $ are holomorphic in $S$, They also behave well when
$\operatorname{Im}\mathcal{D}\rightarrow\pm\infty$, so $\omega\left(
z\right)  =\mathrm{const}$.

\item Fixing the constant by the condition $J_{4}^{\left(  0\right)  }=m^{-2}%
$, we obtain%
\begin{equation}
\omega\left(  z\right)  =m^{-2}-t\left(  0\right)  =\frac{4\pi\theta\left(
2-m^{2}\right)  }{m^{3}\left(  4-m^{2}\right)  ^{3/2}}%
\end{equation}
One can check that unusual $\theta\left(  2-m^{2}\right)  $ in this formula
cancels the discontinuity of $\left(  m^{2}-2\right)  \,_{2}F_{1}\left[
\genfrac{}{}{0pt}{1}{1,2-\mathcal{D}/2}{5/2-\mathcal{D}/2}%
\left\vert m^{2}-\frac{m^{4}}{4}\right.  \right]  $ in $t\left( \mathcal{D}\right)  $. Using the
properties of the hypergeometric function, we
rewrite the final result as%
\begin{equation}
J_{4}^{\left(  \mathcal{D}\right)  }=\frac{\Gamma\left(  1-\frac{\mathcal{D}%
}{2}\right)  \Gamma\left(  2-\frac{\mathcal{D}}{2}\right)  }{\mathcal{D}%
-3}\left(  \,_{2}F_{1}\left[
\genfrac{}{}{0pt}{1}{1,3-\mathcal{D}}{5/2-\mathcal{D}/2}%
\left\vert \frac{m^{2}}{4}\right.  \right]  +m^{\mathcal{D}-2}\,_{2}%
F_{1}\left[
\genfrac{}{}{0pt}{1}{1,2-\mathcal{D}/2}{3/2}%
\left\vert \frac{m^{2}}{4}\right.  \right]  \right)  ,\nonumber
\end{equation}
which coincides with the one obtained in Ref. \cite{DavyTau1993}.
\end{enumerate}

\section{Topologies with several master integrals}

We would like to discuss now the applicability of the described method for the
topologies containing more then one master integral. In this case the
dimensional recurrence relation (\ref{eq:DRR}) should be understood as the
vector equation with $J^{\left(  \mathcal{D}\right)  },R\left(  \mathcal{D}%
\right)  $ being the column vectors and $C\left(  \mathcal{D}\right)  $ the
matrix. As it follows from the theory of difference equations (see, e.g., Ref.
\cite{Laporta2000}), the solution of the homogeneous part of (\ref{eq:DRR})
can be sought for in the form of factorial series and the general form is%
\begin{equation}
J_{0}^{\left(  \mathcal{D}\right)  }=F\left(  \mathcal{D}\right)
\omega\left(  z\right)  ,
\end{equation}
where $F\left(  \mathcal{D}\right)  $ is the matrix of fundamental solutions
and $\omega\left(  z\right)  =\omega\left(  e^{i\pi\mathcal{D}}\right)  $ is a
column vector of arbitrary periodic functions of $\mathcal{D}$ (with period
$2$). So, to proceed in the same way as before, we have to define the summing
factor as $\Sigma\left(  \mathcal{D}\right)  =F^{-1}\left(  \mathcal{D}%
\right)  $. While the elements of the matrix $F\left(  \mathcal{D}\right)  $
have the form of factorial series and their analytical properties can be
analysed in the same way as those of the nonhomogeneous terms in our
consideration, this is not so for the inverse matrix $F^{-1}\left(
\mathcal{D}\right)  $. To overcome this difficulty, we can use the following
transformation. The matrix $F\left(  \mathcal{D}\right)  $ satisfies the
following matrix equation%
\begin{equation}
F\left(  \mathcal{D}-2\right)  =C\left(  \mathcal{D}\right)  F\left(
\mathcal{D}\right)
\end{equation}
Taking the determinant of both sides, we have the equation%
\begin{equation}
f\left(  \mathcal{D}-2\right)  =c\left(  \mathcal{D}\right)  f\left(
\mathcal{D}\right)  ,
\end{equation}
where $f\left(  \mathcal{D}\right)  =\det F\left(  \mathcal{D}\right)  $,
$c\left(  \mathcal{D}\right)  =\det C\left(  \mathcal{D}\right)  $. The
coefficient $c\left(  \mathcal{D}\right)  $ is a rational function of
$\mathcal{D}$ and can be represented as%
\begin{equation}
c\left(  \mathcal{D}\right)  =c\frac{\prod_{i}\left(  a_{i}-\mathcal{D}%
/2\right)  }{\prod_{j}\left(  b_{j}-\mathcal{D}/2\right)  }%
\end{equation}
Therefore, the general form of $f\left(  \mathcal{D}\right)  $ is (cf. with
Eq. (\ref{eq:integrating factor}))%
\begin{equation}
\,f\left(  \mathcal{D}\right)  =c^{-\mathcal{D}/2}\frac{\prod_{i=1}^{n}%
\Gamma\left(  a_{i}-\mathcal{D}/2\right)  }{\prod_{j=1}^{m}\Gamma\left(
b_{j}-\mathcal{D}/2\right)  }\omega_{0}\left(  z\right)
\end{equation}
Now we may use the formula%
\begin{equation}
F^{-1}\left(  \mathcal{D}\right)  =\,f^{-1}\left(  \mathcal{D}\right)
\tilde{F}\left(  \mathcal{D}\right)  ,
\end{equation}
where $\tilde{F}\left(  \mathcal{D}\right)  $ is the adjugate of $F\left(
\mathcal{D}\right)  $. This allows us to choose the summing factor in the form%
\begin{equation}
\Sigma\left(  \mathcal{D}\right)  =c^{\mathcal{D}/2}\frac{\prod_{j=1}%
^{m}\Gamma\left(  b_{j}-\mathcal{D}/2\right)  }{\prod_{i=1}^{n}\Gamma\left(
a_{i}-\mathcal{D}/2\right)  }\tilde{F}\left(  \mathcal{D}\right)
\end{equation}
This formula does not contain the sums in the denominator which allows one to perform the
analysis of its analytical properties in the same manner as we did before. Again, the choice of
the summing factor is not unique due to the possibility of its multiplication from the left by any
periodic matrix $\Omega\left(  z\right)  $. The examples of application of this technique to the
topologies with several master integrals will be given elsewhere.

\section{Conclusions}

We have presented here the approach to the calculation of the master integrals based on
Tarasov's dimensional recurrence relation and on the analytical properties of the loop
integrals as functions of the complex variable $\mathcal{D}$. The results obtained within this
approach have the form of exponentially converging series, in contrast to those obtained with
many other methods having only power-like convergence. Fast convergence of the sums
allow one to apply the \texttt{pslq} algorithm in order to obtain the representation of the
$\epsilon$-expansion of the integral in terms of the conventional transcendental numbers, like
multiple $\zeta$-functions. We have limited the presentation to the cases in which master
integrals of subtopologies are expressed in terms of $\Gamma$-functions. In a more general
case, when the master integrals of subtopologies are expressed via hypergeometric sums
and/or the main topology contains more than one master integral, the analysis of the analytic
properties, though being more complicated, can be performed in the same way. Presumably,
this approach can be applied to the calculation of various master integrals with one scale,
including, in particular, the three-loop onshell vertex integrals not yet calculated.

\begin{ack}
I am grateful to my colleagues G.G. Kirilin, A.V. Pomeransky, and I.S. Terekhov for useful
discussions and for the interest to this work. Special thanks go to G.G. Kirilin for providing
some of his considerations concerning second example in this paper. I appreciate the help of
A.G. Grozin with finding the correct references for first example. I would like to thank also K.G.
Chetyrkin and V.A. Smirnov for informative and stimulating discussions and also for the
interest to the dimensional recurrence method. The third example of this paper stems from the
communications with V.A. Smirnov. This work is supported by RFBR (grant No. 07-02-00953)
and DFG (grant No. GZ436RUS113/769/0-2).
\end{ack}

\bibliographystyle{elsart-num}

\end{document}